\documentclass[journal,onecolumn,draftclsnofoot]{IEEEtran}

\usepackage{xcolor,soul,framed} 
\colorlet{shadecolor}{yellow}
\usepackage[utf8]{inputenc}
\usepackage[pdftex]{graphicx}
\graphicspath{{../pdf/}{../jpeg/}}
\DeclareGraphicsExtensions{.pdf,.jpeg,.png}
\usepackage[cmex10]{amsmath}

\usepackage{array}
\usepackage{mdwmath}
\usepackage{mdwtab}
\usepackage{eqparbox}
\usepackage{url}
\usepackage{booktabs}
\usepackage{stfloats}
\usepackage{subfigure}
\usepackage{cite}
\usepackage{multirow}
\usepackage{amsmath,amsfonts,amssymb}

\def\Vs{V_\mathrm{s}}
\def\VDC{V_\mathrm{DC}}
\def\vVTx{\vec{V}_\mathrm{Tx}}
\def\Xt{X_\mathrm{t}}
\def\RL{R_\mathrm{L}}
\def\T{^\mathrm{T}}
\def\vITx{\vec{I}_\mathrm{Tx}}
\def\vIRp{\vec{I}_\mathrm{Rp}}
\def\IRx{I_\mathrm{Rx}}
\def\vRin{\vec{R}_\mathrm{in}}
\def\vXin{\vec{X}_\mathrm{in}}
\def\Msum{M_\mathrm{sum}}
\def\Pout{P_\mathrm{out}}
\def\RTx{R_\mathrm{Tx}}
\def\RRx{R_\mathrm{Rx}}
\def\RRp{R_\mathrm{Rp}}

\begin{document}
\bstctlcite{IEEEexample:BSTcontrol}
    \title{Omnidirectional Wireless Power Transfer with Automatic Power Flow Control}
  \author{
    Prasad~Jayathurathnage,
    Xiaojie~Dang, 
    Fu~Liu,
    Constantin~Simovski,
    and~Sergei~A.~Tretyakov
}

\markboth{}{}

\maketitle

\begin{abstract}
We present an omnidirectional wireless power transfer (WPT) system capable of automatic power flow control using three orthogonal transmitter (Tx)-repeater (Rp) pairs. The power drawn from each transmitter is automatically adjusted depending on the mutual inductance between the receiver and the Tx-Rp pair. The proposed approach enables the receiver to harvest  almost uniform power with high efficiency ($90\%$) regardless of its position. 
\end{abstract}

\IEEEpeerreviewmaketitle

\section{Introduction}

\IEEEPARstart{W}{ireless} power transfer (WPT) is increasingly becoming a popular technique for providing power to electronic devices. Although it is convenient as no wire plugging is needed, misalignment between the receiver (Rx) coil and transmitter (Tx) coil reduce the transferred power and efficiency. To overcome this misalignment restriction, researchers have proposed several approaches, among which omnidirectional WPT with multiple three-dimensional (3D) coils is a prominent one as it provides more freedom for the position of the receivers. 
For example, a Tx with multiple orthogonal windings was proposed to enable multi-angle WPT~\cite{wang2012enabling,jonah2013orientation}. However, since the total magnetic field is the vector sum of the magnetic fields induced by all the coils, the total magnetic field is fixed at a particular direction, therefore, omnidirectional WPT is impossible to achieve. 
 
To solve this problem, a nonidentical current control technique is employed in 3D transmitters to steer the direction of synthetic magnetic field depending on the position of the Rx \cite{2014_Hui_2D-3D,2015_Hui_omni_control}. There are several controlling methods of the Tx currents such as amplitude modulation, phase shifting, and frequency modulation. The amplitude modulation, where the amplitude of the individual Tx current is controlled, can direct the total magnetic field vector toward the Rx, enabling \emph{directional WPT} \cite{2016_Hui_mathanalysis,2015_Hui_omni_control}. In this scenario, the optimal magnitude of a Tx current is proportional to the mutual inductance between that Tx and the Rx, and the power received by the load is proportional to the square sum of all the mutual inductances between each Tx and the Rx  \cite{2016_Hui_mathanalysis}. A prerequisite of this action is the knowledge of all the mutual couplings, that is, the knowledge of the Rx position. For example, Zhang et al. in \cite{2015_Hui_omni_control} present a control method to detect the Rx position and focus the power flow toward the targeted receiver. However, this approach requires complex and high-precision control methods to guarantee an efficient directional WPT. 
On the other hand, the phase shifting method, where the phases of Tx currents are controlled, can create a rotating magnetic field around the 3D-transmitter, enabling \emph{rotational WPT}. For example, a $120^{\circ}$ phase difference between the currents of the three orthogonal Txs is used to create a 3D rotating magnetic field in~\cite{2014_Hui_2D-3D}. However, low transfer efficiency is inevitable in this scenario because power transfer is only effective when the field vector is oriented towards the Rx.
 
\begin{figure}[!b]
 \begin{center}
 \vspace{-5mm}
 \includegraphics[width=0.5\columnwidth]{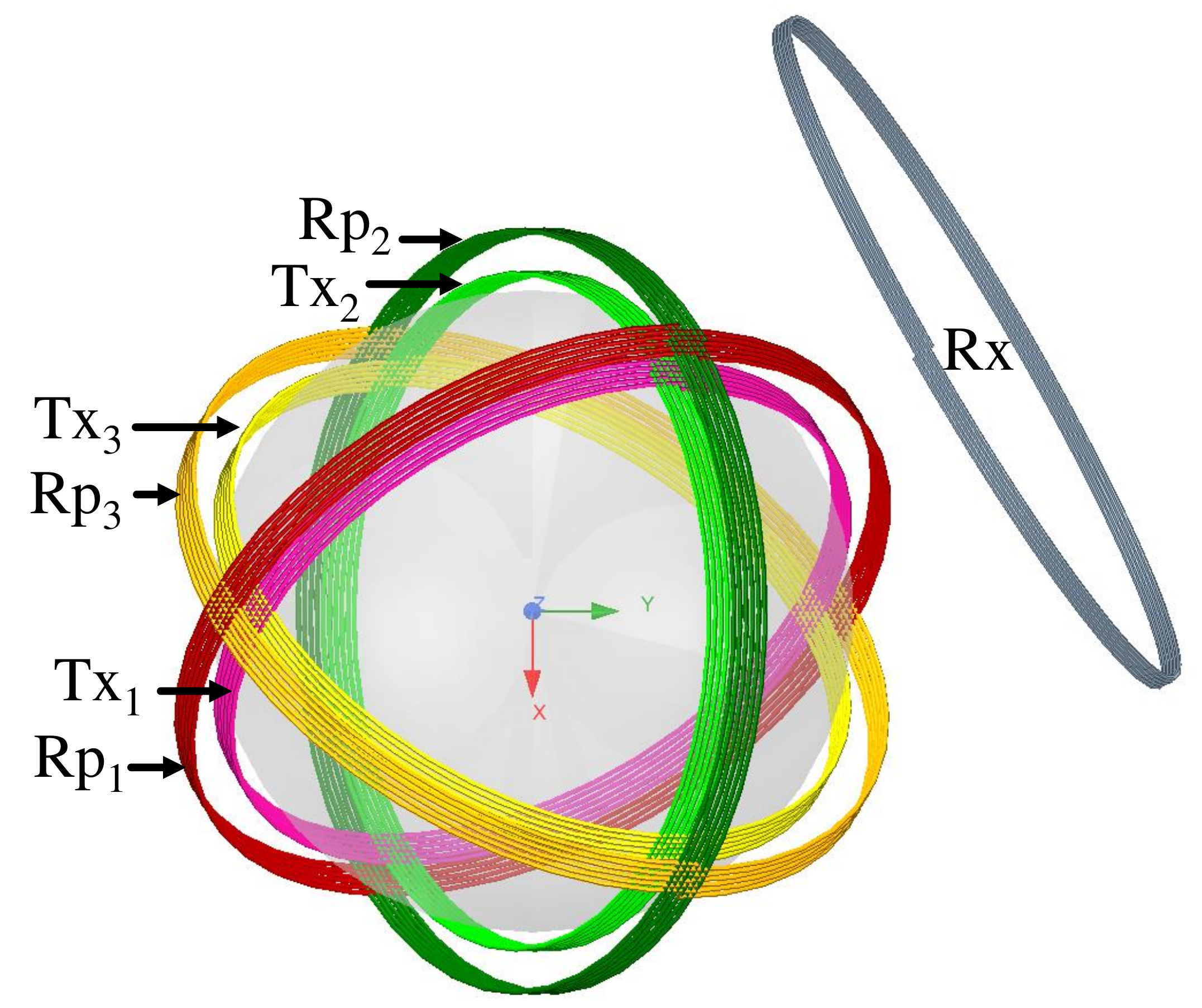}
 \caption{The proposed omnidirectional WPT system with three orthogonal Tx-Rp pairs}\label{Fig.Omnidirectional_WP_coils}
 \end{center}
\end{figure}

In this paper, we propose and experimentally verify a novel omnidirectional WPT system to realize the performance equivalent to \emph{directional WPT} with automatic power flow control. The proposed WPT system consists of three transmitter-repeater pairs, which are placed in an orthogonal way and form three independent power channels, as illustrated in Fig.~\ref{Fig.Omnidirectional_WP_coils}. To achieve efficient omnidirectional WPT operation, we propose a setup for automatic tuning the amplitude of each Tx current to be proportional to the mutual inductance between its repeater (Rp) and the Rx, which is similar to the optimal current distribution in the directional WPT scenario. Only a simple control is needed in the transmitter side to keep Tx coil current in-phase with the supply voltage to ensure the constructive mutual coupling. The experimental results of the laboratory prototype validate the omnidirectional WPT performance of the proposed system. 

\section{Theoretical Analysis}

\subsection{The Equivalent Circuit Analysis}
\begin{figure}[!b]
 \begin{center}
 \includegraphics[width=0.6\columnwidth]{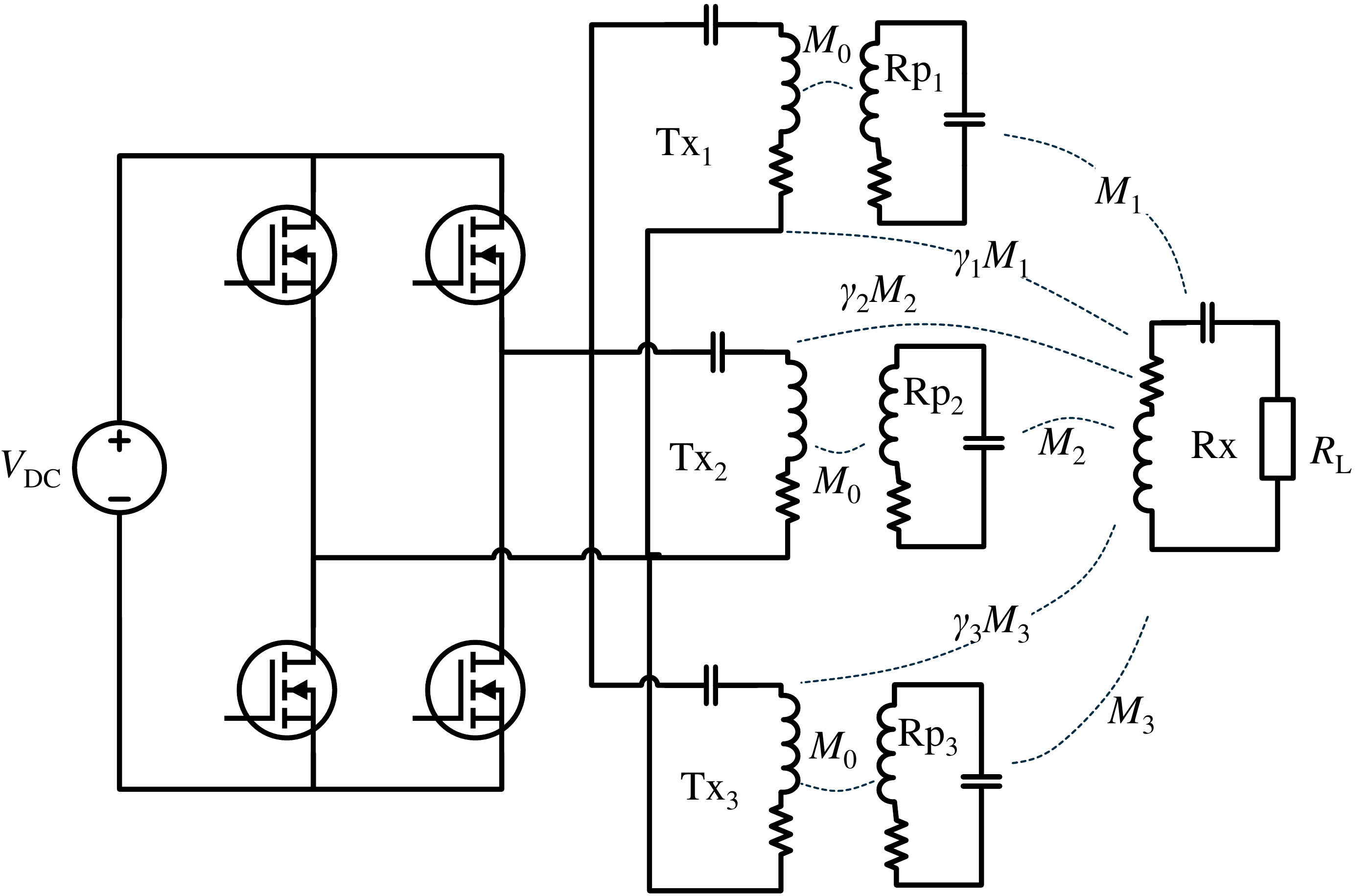}
 \caption{The equivalent circuit of the proposed WPT system with three Tx-Rp pairs and the Rx}\label{Fig.EEq_cct}
 \vspace{-7mm}
 \end{center}
\end{figure}
The equivalent circuit of the proposed omnidirectional WPT system is shown in Fig.~\ref{Fig.EEq_cct}. All the Txs are series compensated and connected in parallel to the output of a full-bridge inverter. Because of the high-Q resonance network, we assume that the three Txs are connected to an ideal sinusoidal voltage source with an RMS voltage of $\Vs=2\sqrt{2}\VDC/\pi$, where $\VDC$ is the DC source voltage. As the three Tx-Rp pairs are orthogonal to each other, the cross couplings between the transmitter coils and repeater coils that belong to different pairs are negligible, which ensures that the Tx-Rp pairs form three independent power channels. The master equation of this WPT system can be written as
\begin{equation}
	\begin{bmatrix}
		\vVTx \\ \vec{0} \\ 0
	\end{bmatrix}=
	\begin{bmatrix}
		\RTx+j \Xt						& j \omega_0 M_0 		& j \omega_0 \vec{\Gamma}\\
		j \omega_0 M_0 				& \RRp 					& j \omega_0 \vec{M}\\
		j \omega_0 \vec{\Gamma}\T 	& j \omega_0 \vec{M}\T 	& \RL+\RRx
	\end{bmatrix}
	\begin{bmatrix}
		\vITx \\ \vIRp \\ \IRx
	\end{bmatrix},
\end{equation}
where the left (right) column vector represents the voltages (currents) in the Txs, Rps, and Rx, with $\vVTx=\left[\Vs \; \Vs \; \Vs \right]\T$ denoting the identical source voltages, $\vITx=\left[I_\mathrm{Tx1} \; I_\mathrm{Tx2} \; I_\mathrm{Tx3} \right]\T$, $\vIRp=\left[I_\mathrm{Rp1} \; I_\mathrm{Rp2} \; I_\mathrm{Rp3} \right]\T$, and $\IRx$ denoting the currents in the Tx, Rp, and Rx. The impedance matrix relates the voltages and currents of the whole system, with $\RTx, \RRp$, and $\RRx$  being the parasitic resistances of Tx, Rp, and Rx coils, respectively. $R_L$ is load resistance, $M_0$ is the mutual inductance between the transmitter and repeater belonging to the same channel, $\vec{M}=\left[M_1 \; M_2 \; M_3 \right]\T$ denoting the mutual inductance between each repeater coil and the receiver coil and $\vec{\Gamma}=\left[\gamma_1 M_1 \; \gamma_2 M_2 \; \gamma_3 M_3 \right]\T$ denoting the mutual inductance between each transmitter coil and the receiver coil, see~Fig.~\ref{Fig.EEq_cct} for reference. We note that in developing this master equation we have used the same resonant frequency $\omega_0$ (also the working frequency) in the repeater and receiver, while the transmitter has non-zero reactance $\Xt$.

If we assume ideal coils and neglect the parasitic resistances, i.e., $\RTx=\RRp=\RRx \approx 0$, we find that the input resistance $\vRin$ and input reactance $\vXin$ seen from the transmitters are
\begin{eqnarray}
	\vRin & = & \dfrac{M_0^2 \RL}{\Msum \vec{M}}, \\
	\vXin & = & \dfrac{\Xt \vec{M}\cdot\vec{M} - 2\omega_0 M_0 \vec{\Gamma}\cdot\vec{M}}{\Msum \vec{M}},
\end{eqnarray}
where $\Msum=M_1+M_2+M_3$. As all the Tx-Rp pairs are identical, it is reasonable to assume $\gamma_1=\gamma_2=\gamma_3=\gamma$ which implies $\vec{\Gamma}=\gamma\vec{M}$. In this case, all the input reactances can be easily nullified at the same time by choosing
\begin{equation}
\label{Eq.Xt}
	\Xt=2\omega_0 \gamma M_0,
\end{equation}
which is completely independent from receiver position. Finally, under the above conditions, we find the currents in the transmitter, repeater, and receiver as

\begin{eqnarray}
    \label{Eq.ITx}
	\vITx & = & \dfrac{\Msum\vec{M}}{M_0^2}\dfrac{\Vs}{\RL}, \\
	\label{Eq.IRp}
	\vIRp & = & -\gamma \vITx-\dfrac{j\Vs}{\omega_0 M_0}, \\
	\label{Eq.IRx}
	\IRx  & = & -\dfrac{\Msum}{M_0}\dfrac{\Vs}{\RL}.
\end{eqnarray}
We note that Tx currents are proportional to $\vec{M}$, which is analogous to the optimal Tx currents in \emph{directional WPT} \cite{2016_Hui_mathanalysis}. This means that the current in a particular Tx increases with the increase of the mutual coupling between its repeater and the Rx, while it is suppressed when the coupling is very small. This is in fact an essential and desired criterion for high efficiency omnidirectional WPT, because the power transfer contribution from each Tx is automatically adjusted based on the mutual inductance between the receiver and the corresponding repeater. This automatic power flow control capability is also reflected in the Rx current $\IRx$. As $\IRx$ is proportional to $\Msum$, stable output power can be achieved as long as the total coupling inductance $\Msum$ between the repeaters and receiver is stable.

\subsection{Transmitter Switching Strategy}
When the receiver moves, a good strategy to achieve stable output power is to always ensure constructive magnetic field flux through the receiver from all Tx coils. However, this is impossible without control. For example, let us consider the case shown in Fig.~\ref{Fig.Omnidirectional_WP_coils}. When the receiver moves in the $x-y$ plane while facing the center of the Tx coils, the three mutual inductances composing the vector $\vec{M}$ experience variation between positive and negative values and the variations have 120$^{\circ}$ phase difference. As a result, the total magnetic flux is not always added in phase to feed the receiver. This will decrease $\Msum$ and reduce the transferred power.
In order to resolve this issue, we can simply employ a Tx-side control strategy, inverting the direction of the Rp current by reversing the terminals of that transmitter which has negative  mutual inductance with the receiver. This inversion will change the sign of the effective mutual inductance and the inductive coupling becomes constructive. From Eq.~\eqref{Eq.ITx}, we see that the detection of the destructive mutual coupling can be done by measuring the phase of all three Tx currents separately. Thus, the Tx-side control strategy allows us to keep Tx current and its terminal voltage in-phase, no any feedback from the Rx-side is needed. It only requires additional switches at the Tx terminals or three separate power converters connected to each Tx. Using this strategy, the total mutual coupling $\Msum$ is redefined as $\Msum=|M_1|+|M_2|+|M_3|$. High efficiency and optimal output power can be then achieved regardless of the receiver position. 

\subsection{WPT Performances}
Let us define the performance indicators of the WPT system including the output power at the load and the efficiency using \eqref{Eq.ITx}-\eqref{Eq.IRx}.
The output power $\Pout$ can be calculated as
\begin{equation} \label{Eq.Pout}
	\Pout=\left|\IRx\right|^2 \RL=\left(\dfrac{\Msum}{M_0}\right)^2\dfrac{\Vs^2}{\RL}.
\end{equation}
We can observe that it is proportional to $\Msum^2$, which is comparable to the $\Pout$ profile in the amplitudes-controlled \emph{directional WPT} \cite{2015_Hui_omni_control,2016_Hui_mathanalysis}. 
Next, power efficiency can be calculated as

\begin{eqnarray}
\label{Eq.Efficiency}
    \eta & = & \dfrac{1}{1+\xi_\mathrm{Tx}+\xi_\mathrm{Rp}+\xi_\mathrm{Rx}},\\
 \text{where~~~~~}	\xi_\mathrm{Tx} & = & \dfrac{\vec{M}\cdot\vec{M}}{M_0^2}\dfrac{\RTx}{\RL},~~~~~\xi_\mathrm{Rx}  =  \dfrac{\RRx}{\RL}, \nonumber\\
    \xi_\mathrm{Rp} & = & \dfrac{3 \RRp \RL}{\omega_0^2 \Msum^2} +\dfrac{\gamma^2\vec{M}\cdot\vec{M}}{M_0^2}\dfrac{\RRp}{\RL}\nonumber.
\end{eqnarray}
\noindent The terms $\xi_\mathrm{Tx}$, $\xi_\mathrm{Rp}$, and $\xi_\mathrm{Rx}$ represent the loss ratios between the losses in Txs, Rps, and Rx and the output power, respectively. The term $\xi_\mathrm{Rx}$ is very small as the coil resistance $\RRx$ is usually much smaller than the load resistance. For the same reason, the losses in the Txs are very small and can be further minimized by increasing the coupling $M_0$. Therefore, the most prominent losses are from the Rps. To improve the efficiency, the coil resistance $\RRp$ and $\gamma$ should be as low as possible while having high $\Msum^2$. 

In summary, the proposed omnidirectional WPT transmitter provides stable output power with high efficiency nearly equivalent to that of the optimal \emph{directional WPT}. Only a simple control at the Tx side is needed to deal with destructive mutual coupling (which is inherent also to other omnidirectional WPT systems).

\section{Experimental Validation}
\subsection{Description of the WPT Setup}
The proposed concept of the omnidirectional WPT system is verified using an experimental prototype operating at 593~kHz. Litz wire is used for the coils to minimize the parasitic resistances and it is winded in helical shape for all the coils. To fulfill the desired condition $\gamma_1 \approx \gamma_2 \approx\gamma_3 < 1$ for all the Rx positions, the diameter of the Tx coils is chosen to be $260$~mm, slightly smaller than that of the Rp coils, which is $300$~mm. In addition, the Tx and Rp coils have $3$ turns. On the other hand, the Rx coil has $10$ turns with diameter of $300$~mm. The measured electrical parameters of the WPT coils are listed in Table~\ref{Table.coil_measure}.

\begin{table}[!b]
\caption{Measured Parameters of the WPT Coils}
\centering
\label{Table.coil_measure}
\begin{tabular}{@{}lcccc@{}}
\toprule\toprule
  &  \multicolumn{1}{l}{Inductance }&  \multicolumn{1}{l}{Resistance} & \multicolumn{1}{l}{$Q$} & \multicolumn{1}{l}{Resonance} \\ 
                      &($\mu$H) & (m$\Omega$)     &     & (kHz)\\
  \midrule
           Tx1 (Red)      &6.41   & 49     & 490   & 312.7          \\
           Tx2 (Green)    &6.33   & 47     & 505   & 310          \\
           Tx3 (Yellow)   &6.43   & 39     & 620   & 316.3         \\
           Rp1 (Red)      &7.39   & 55     & 510   & 592.8         \\
           Rp2 (Green)    &7.45   & 55     & 512   & 593.7         \\
           Rp3 (Yellow)   &7.53   & 37     & 765   & 592.3         \\
           Rx            &75.93  & 469    & 610   & 592.6         \\
\bottomrule\bottomrule
\end{tabular}
\end{table}

\begin{figure}[!tb]
 \begin{center}
 \includegraphics[width=0.6\columnwidth]{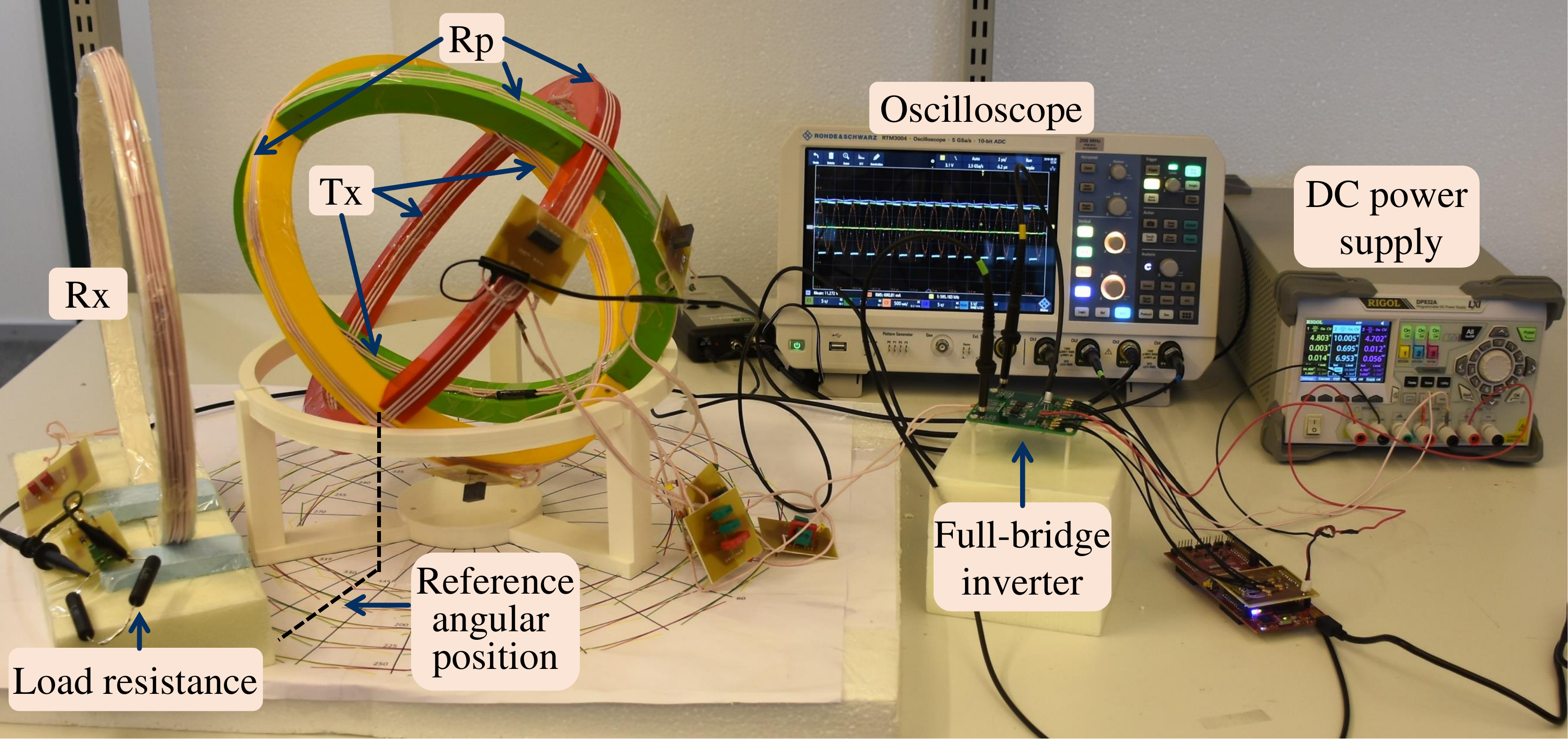}
 \caption{The experimental setup.}\label{Fig.The_experimental_setup}
  \vspace{-5mm}
 \end{center}
\end{figure}

In the experimental setup, the transmitter is placed so that all three Tx (and Rp) coils make the same inclination angle with the horizontal plane, as illustrated in Fig.~\ref{Fig.The_experimental_setup}. For visual aid, the coils of the three transmitters Tx1, Tx2, and Tx3 (and the repeaters Rp1, Rp2, and Rp3) are winded in support structures with color red, green, and yellow, respectively. The Rx is vertically placed at the same level and moves around the Txs in the horizontal plane with fixed distance $200$~mm from the center of Txs. Therefore, the Rx position can be denoted by a rotation angle and we choose the zero angle reference to be the position where the Rx is facing Tx2, as indicated in Fig.~\ref{Fig.The_experimental_setup}. For the power source, a full-bridge inverter is built using two Gallium Nitride half-bridges ${\rm LMG5200}$. The load resistance is connected to the Rx coil through a diode rectifier. As the primary focus of this letter is to introduce the omnidirectional WPT concept, the transmitter terminals are manually switched to deal with negative mutual inductance. 

\subsection {Results and Discussions}

\begin{figure}[!b]
 \begin{center}
 \includegraphics[width=0.7\columnwidth]{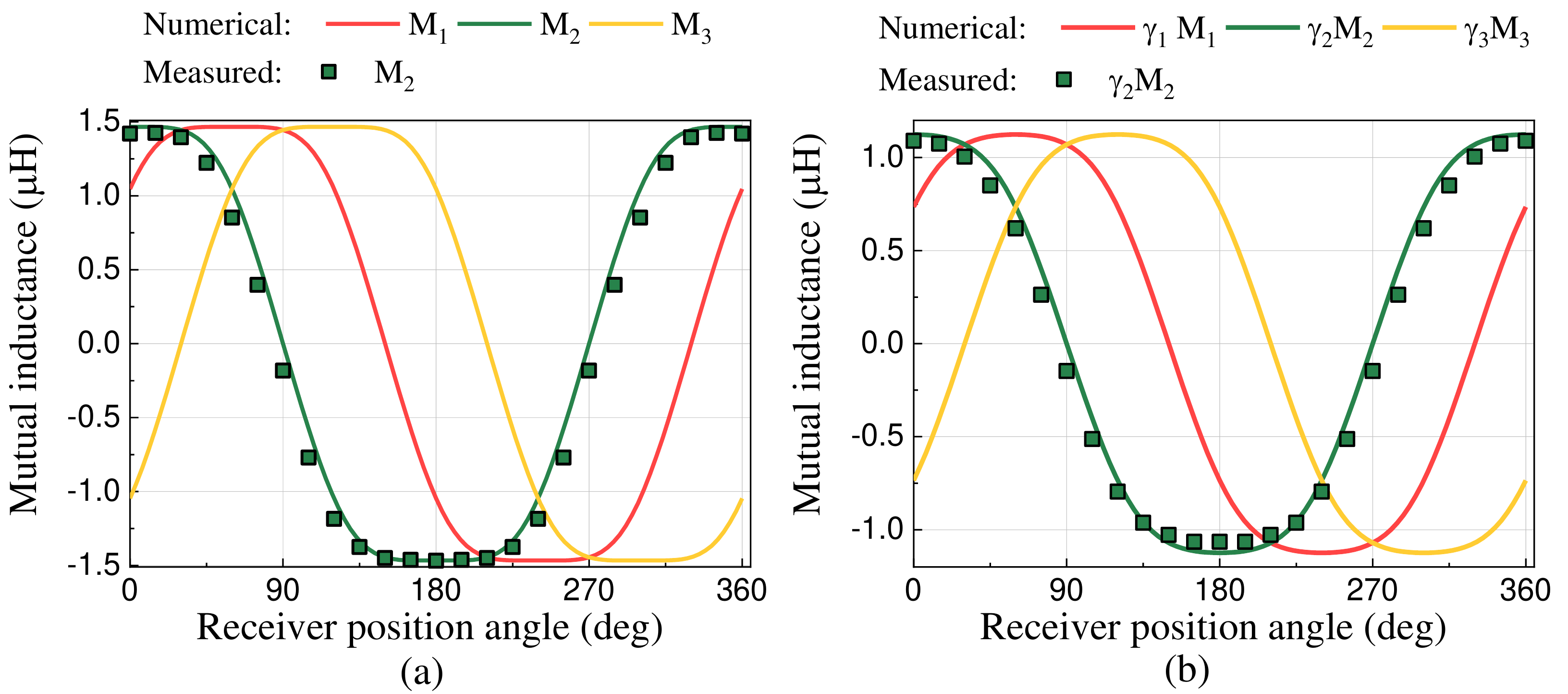}
  \vspace{-5mm}
 \caption{Mutual inductance variations with respect to the receiver position angle at $200$~mm distance.}\label{Fig.M_variations_vs_angle_200mm}
 \end{center}
\end{figure}

First, the mutual inductances between Txs-Rx and Rps-Rx are numerically calculated using the approach in \cite{babic2010mutual} and experimentally verified. The results are presented in Fig.~\ref{Fig.M_variations_vs_angle_200mm}. It is clear that the mutual couplings experience positive/negative value variations with $120^{\circ}$ shift, and the measured $\gamma$ varies between $0.66$ and $0.75$. The measured mutual inductance $M_0$ is about $3.1~{\rm \mu H}$, while the cross coupling between the coils that belong to different Tx-Rp pairs is around $20$~nH, much smaller than the mutual inductances between Rp and Rx. Therefore, the claim of completely uncoupled and independent Tx-Rp channels is well justified. Using \eqref{Eq.Xt} and the measured $M_0$ and $\gamma$ values, we obtain that the required value for the reactive impedance $\Xt$ of Txs is between $15.2~{\rm \Omega}$ and $17.3~{\rm \Omega}$. In fact, slightly inductive input impedance is preferred to achieve soft-switching of the converter \cite{Pantic_ZCS}. Therefore, $\Xt$ is set to be $17.5~{\rm \Omega}$ at the working frequency in the experiment by properly tuning the resonance frequencies of the Tx coils. On the other hand, all the repeaters and the receiver are resonant at the working frequency, and the measured resonant frequencies are also listed in Table~\ref{Table.coil_measure}.

\begin{figure}[!t]
 \begin{center}
 \includegraphics[width=0.7\columnwidth]{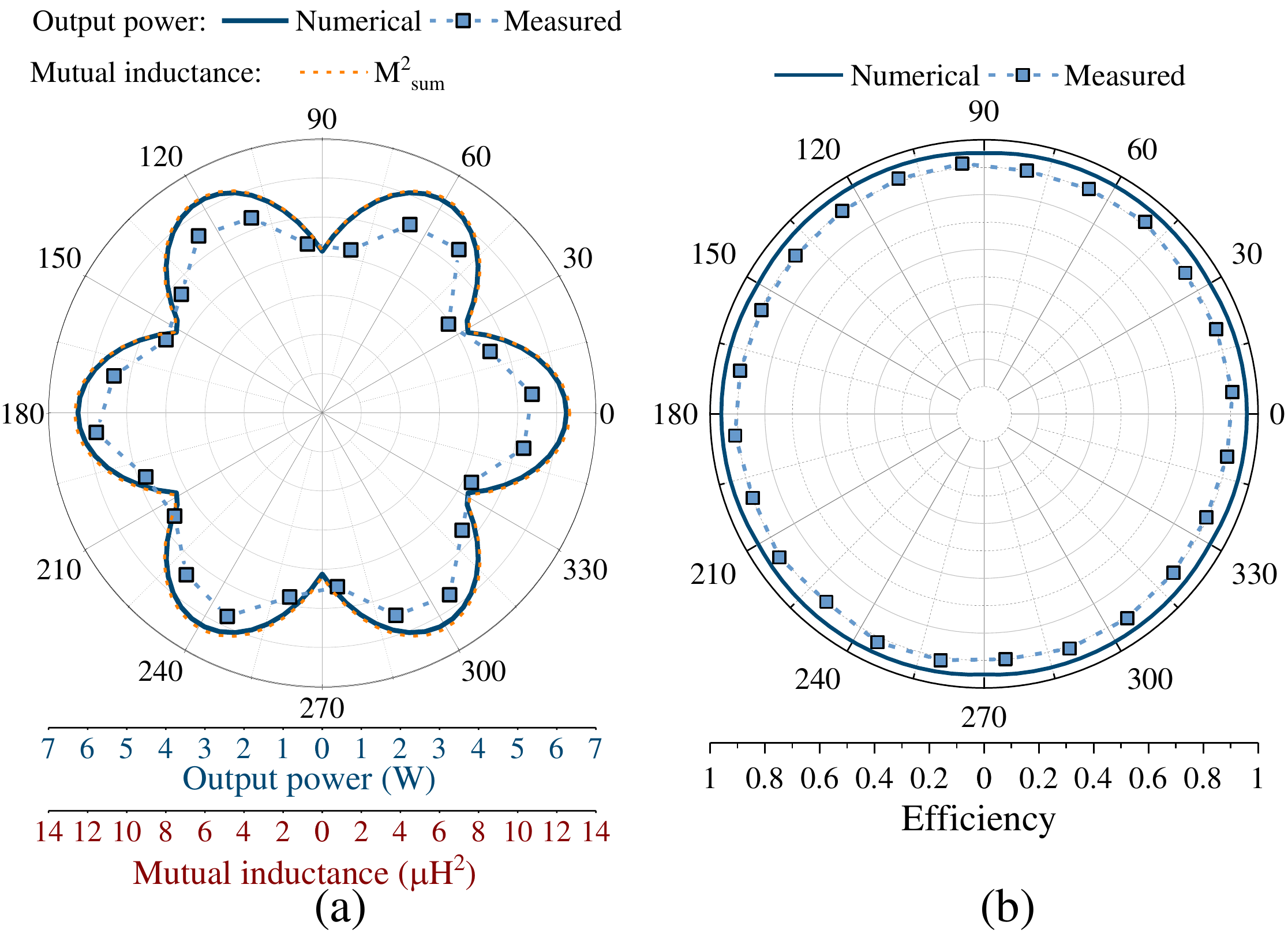}
 \vspace{-6mm}
 \caption{Comparison of numerical and measured results with respect to receiver position angle at $200$~mm transfer distance; (a)~Output power and mutual inductance $\Msum^2$; Output power is measured at the DC load of $20{\rm \Omega}$, and (b)~Efficiency; Efficiency is measured from the DC power source to the DC load. } \vspace{-8mm}
 \label{Fig.Experimental_Pout_eta}
 \end{center}
\end{figure}

Next, the performances of the omnidirectional WPT system is experimentally verified when the Rx is circulating around the transmitter. The measured output power and efficiency are illustrated in Fig.~\ref{Fig.Experimental_Pout_eta}. As we can observe, the output power is quite stable while varying within an acceptable range when the the receiver position varies for $360^{\circ}$. It is worth to note that the output power profile overlaps with $\Msum^2$, validating the theoretical derivation in \eqref{Eq.Pout}. The six dips at $30^{\circ}+n60^{\circ} (n=0,1,...,5)$ come from the minimum $\Msum$ values when the receiver is perpendicular to one of the three Tx coils and only two Txs contribute to the power transfer. Furthermore, from Fig.~\ref{Fig.Experimental_Pout_eta}(b) we can observe that the measured DC-to-DC efficiency is almost constant at around $90\%$ regardless of Rx position. The slightly lower measured efficiency (about $5\%$ less compared to the numerical one) is due to the fact that only coil losses are considered in the numerical calculations while the measured one includes also the losses in the full-bridge inverter and the rectifier. In fact, the experimental performance is in excellent agreement with the theoretical results.

\begin{figure}[!t]
 \begin{center}
 \includegraphics[width=0.7\columnwidth]{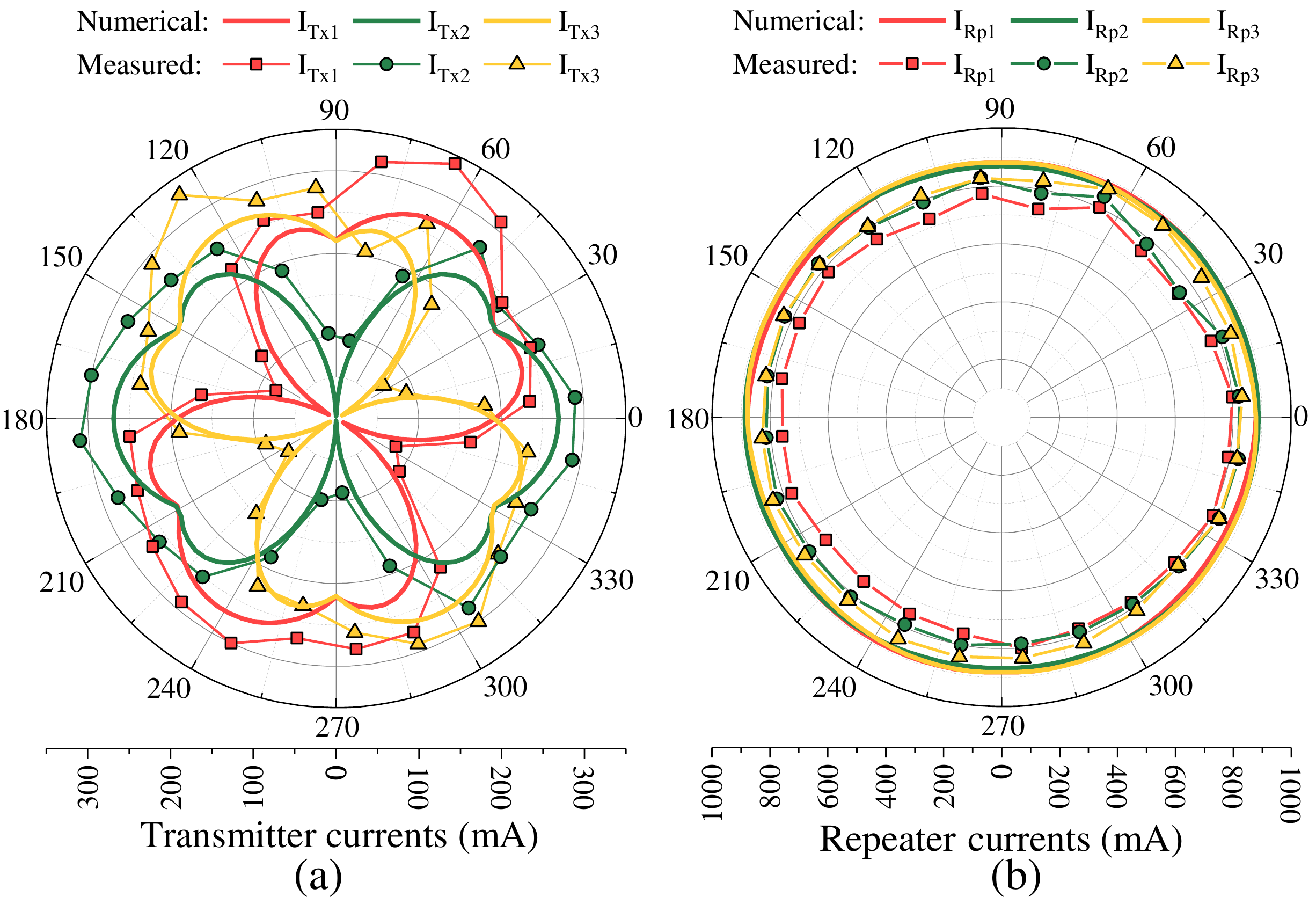}
 \caption{Comparison of numerical and measured currents with respect to receiver position angle at $200$~mm distance; (a)~Transmitter currents, and (b)~Repeater currents. }
 \label{Fig.Txcurrent_vs_angle_200mm}
 \vspace{-5mm}
 \end{center}
\end{figure}

Finally, the measured Tx and Rp currents are presented in Fig.~\ref{Fig.Txcurrent_vs_angle_200mm} to explain the basic principle of the proposed automatic power flow control. While the variations of Tx currents follow the mutual inductance profile [see Figs.~\ref{Fig.M_variations_vs_angle_200mm} and~\ref{Fig.Txcurrent_vs_angle_200mm}(a)], the Rp currents remain almost constant against the rotational angle of receiver [see Fig.~\ref{Fig.Txcurrent_vs_angle_200mm}(b)]. These results are consistent with the theoretical results presented in \eqref{Eq.ITx} and \eqref{Eq.IRx}, as we expected.

\section{Conclusion}
In this paper, we have proposed a novel omnidirectional WPT system with automatic power flow control capability using three pairs of orthogonal transmitters and repeaters as three independent power channels. It has been theoretically and experimentally confirmed that the proposed WPT system can generate true omnidirectional WPT with high efficiency and stable output power. Tx coil currents are proportional to the mutual inductance between each repeater and the receiver, which guarantees automatic power control depending on the receiver position. Compared to the conventional WPT systems with complex control, the proposed system only requires switching Tx-terminals to ensure in-phase Tx currents and their terminal voltages. This simple tuning can be implemented using only Tx-side sensors, that ensures constructive power contribution from each channel in spite of destructive mutual coupling. The experimental system features power transfer efficiency around $90\%$ over the full range of receiver positions at transfer distance of $200$~mm.

\ifCLASSOPTIONcaptionsoff
 \newpage
\fi

\bibliographystyle{IEEEtran}
\bibliography{IEEEabrv,Bibliography}

\vfill
\end{document}